\documentclass[12pt]{article}

\usepackage[dvips]{graphicx}

\begin{document}

\begin{titlepage}

\title{Light Pair Correction\break to Bhabha Scattering at Small Angle}

\author{G.Montagna $^a$ \and M.Moretti $^{b,c}$ \and O.Nicrosini $^a$ \and
	A.Pallavicini $^a$ \and F.Piccinini $^a$}

\date{\vspace*{-2em}}

\maketitle

\noindent
{\footnotesize\it
$^a$ Dipartimento di Fisica Nucleare e Teorica - Universit\`a di Pavia and
     INFN - Sezione di Pavia, via A. Bassi 6, Pavia, Italy}\\
{\footnotesize\it
$^b$ Theory Division, CERN,  CH-1211 Geneva 23, Switzerland}\\
{\footnotesize\it
$^c$ Dipartimento di Fisica - Universit\`a di Ferrara and
     INFN - Sezione di Ferrara, Ferrara, Italy}

\begin{abstract}
		This work deals with the computation of electron pair
		correction to small angle Bhabha scattering, in order
		to contribute to the improvement of luminometry
		precision at {\sc lep/slc} below $0.1\%$ theoretical accuracy.

		The exact {\sc qed} four-fermion matrix element for
		$e^+e^-\rightarrow e^+e^-e^+e^-$, including all
		diagrams and mass terms, is computed and different
		Feynman graph topologies are studied to quantify
		the error of approximate calculations present in the
		literature.

		Several numerical results, obtained by a Monte
		Carlo program with full matrix element, initial-state
		radiation via collinear structure functions, and
		realistic event selections, are shown and critically
		compared with the existing ones.

		The present calculation, together with recent progress
		in the sector of $O(\alpha^2)$ purely photonic corrections,
		contributes to achieve a total theoretical error in
		luminometry at the $0.05\%$ level, close to the current
		experimental precision and important in view of the final
		analysis of the electroweak precision data.
\end{abstract}

\begin{quote}
		{\sc pacs}: 02.70.Lq,12.15.Lk,13.40.Ks,13.85.Hd   
\end{quote}

\newlength{\ppheight}
\setlength{\ppheight}{\vsize}
\vspace*{-\ppheight}
\rightline{\tt FNT/T-98/11}
\rightline{\tt CERN-TH/98-356}

\end{titlepage}

\section{\label{intro} Introduction}

The high-precision determination of the machine luminosity at
{\sc lep/slc} is an essential ingredient of the success of 
precision tests of the electroweak interactions on top of the $Z$
resonance \cite{review}.

As well known, the Bhabha scattering process at small angle (of the
order of a few degrees) is the reference reaction used for luminosity 
monitoring at {\sc lep/slc}, owing to its large cross section (dominated by
$t$-channel photon exchange) and its substantial independence of
purely electroweak effects. Experimental efforts in the development of 
efficient, dedicated luminometry detectors, as well as precision
calculations of the small-angle Bhabha (hereafter {\sc sabh}) scattering cross
section both contribute to achieve a measurement of the ``$Z$ factories'' luminosity 
with a total relative error at the $0.1\%$ level \cite{review,exp,common}.
On the experimental side, the present total uncertainty is smaller than
$0.1\%$ \cite{exp}, close to the $0.05$ level \cite{ward}. As far as the theory
contribution to the luminosity measurement is concerned, the
estimate of the theoretical errors, used by the {\sc lep} collaborations,
is summarized in table \ref{sabs} \cite{common} for centre of mass
energies around and above the $Z$ resonance.

\begin{table}[ht]
\caption[sabs]{\label{sabs}
               Theoretical error in {\sc sabh} scattering according
               to ref.~\cite{common} at typical {\sc lep1} and {\sc lep2}
               energies.}
\medskip
\begin{center}
\begin{tabular}{|l||c|c|} \hline
Type of correction/error & {\sc lep1} ($\%$) & {\sc lep2} ($\%$)\\ \hline \hline
missing photonic $O(\alpha^2L)$   & $0.100 $ & $0.200$ \\
missing photonic $O(\alpha^3L^3)$ & $0.015 $ & $0.030$ \\
vacuum polarization      & $0.040 $ & $0.100$ \\
light pairs              & $0.030 $ & $0.050$ \\
$Z$-exchange             & $0.015 $ & $0.000$ \\ \hline
total                    & $0.110 $ & $0.250$ \\ \hline
\end{tabular}
\end{center}
\end{table}

Some comments on table \ref{sabs} are in order. The components of the theoretical
error refer to the {\sc sabh} scattering cross section, for any typical event
selection of {\sc lep} experiments, as computed by the program {\tt
BHLUMI v4.03} \cite{bhl}.
The largely dominating source of theoretical error 
is due to the missing part of $O(\alpha^2 L)$ subleading photonic corrections,
where $L = \ln(-t/m^2)$ is the collinear logarithm in $t$-channel scattering.
Also the contribution of the missing part of the leading $O(\alpha^3L^3)$
corrections is of photonic nature. The vacuum polarization entry
is the effect of the uncertainty in the hadronic contribution
to the running of $\alpha_{\rm QED}$, when considering the parameterization and
relative error estimate of ref.~\cite{oldvacuum}.
The next contribution is the uncertainty introduced by the corrections due
to the production of light pairs, chiefly $e^+ e^-$ ones. The last
entry refers to the uncertainty associated to the treatment of the
$\gamma$-$Z$ interference.
More details about the strategy adopted in order to estimate the various 
sources of theoretical error can be found in ref.~\cite{common}.

After the analysis of ref. \cite{common}, important theoretical
developments took place. Additional work in the sector of two-loop
photonic corrections \cite{pv,kr} led to the conclusion that the
perturbative contribution due to the uncontrolled part of $O(\alpha^2L)$
corrections does not exceed the $0.03\%$ level. This conclusion has been 
very recently reinforced by the detailed analysis of ref.~\cite{ward}. Furthermore,
new determinations \cite{vacuum} of $\alpha_{\rm QED}$ lower the error
on hadronic contribution to vacuum polarization in $s$-channel
processes at $\sqrt{s}=M_Z$. This might affect {\sc sabh}
scattering too, although no dedicated analysis for the low-angle 
regime exists yet.

As a consequence of this progress, it is relevant
to reduce the uncertainty associated to the light pair
contribution. At present, the calculations available in the
literature and used to estimate the light pair uncertainty
as given in table \ref{sabs} concern a Monte Carlo (hereafter
{\sc mc}) computation based on an approximate $t$-channel matrix
element \cite{jadach} and to an analytical approach with fixed
event selections \cite{russi}. Previous leading logarithmic
evaluations of the dominant light pair contribution to {\sc sabh}
can be found in ref.~\cite{llog}.

In order to improve the existing situation and 
contribute to the lowering of the light pair error, in 
the present paper a {\sc mc} calculation is drawn with the
exact $e^+ e^- \to e^+ e^-e^+ e^-$ matrix element and taking 
into account realistic event selections. The analysis is chiefly
presented at {\sc lep1/slc} energies ($\sqrt{s}=92{\rm~GeV}$),
but numerical results are shown at {\sc lep2} energies
($\sqrt{s}=176{\rm~GeV}$) too.
The impact of the present calculation in the reduction of
the theoretical error for {\sc lep/slc} luminosity measurement 
is also discussed.

The outline of the paper is as follows. The details concerning the treatment of 
phase space are described in section \ref{phsp}, while in section \ref{caloes}
the selection criteria considered in the present study are reviewed.
Section \ref{algdyn} is devoted to describe the calculation.
The last sections contain a discussion of numerical 
results, including comparison with existing analytical calculations 
(section \ref{ancomp}) and approximate {\sc mc} results (section \ref{tchn}),
as well as study of the effect of initial-state radiation (hereafter {\sc isr})
(section \ref{isrcorr}). The conclusions and possible developments are
given in section \ref{end}.

\section{\label{phsp} Phase Space Generation}

Pair corrections to Bhabha scattering lead to a four-body kinematics and to
an 8-dimensional phase space. Expansion in few body processes can greatly
simplify the phase space parameterization. The choice between different equivalent
expansions should be suggested by the relevant dynamics.

\begin{figure}[ht]
\begin{center}
\includegraphics[bb=140 600 254 681,scale=1.]{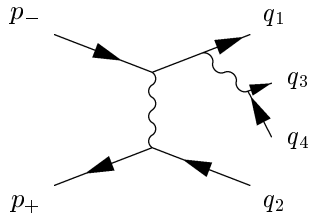}
\end{center}
\caption[bremsstrahlung]{\label{tch}
                         One of the sixteen bremsstrahlung graphs representing
                         the leading $t$-channel dynamics.}
\end{figure}

At high energies and small momentum transfer, which is of interest for
luminosity measurements at {\sc lep}, the leading contribution 
to cross section is given by bremsstrahlung Feynman graphs sketched
in figure \ref{tch}, i.e. by $t$-channel photon exchange dynamics.
Indeed, since the relevant dynamics is dominated by pure {\sc qed} processes,
weak effects, such as $Z$-exchange or multiperipheral graphs mediated
by at least one $Z$ boson, will be neglected in the following.
Further the present study deals mainly with electron pair production,
because it is known \cite{common,russi} that heavier particles give a
much smaller contribution.\\

Let us define the photon momentum $k\equiv q_3+q_4$ and its energy
by $\omega\equiv k_0$.
The core of bremsstrahlung contribution is
given by the soft pair approximation, i.e. the limit
$|t|\gg \omega,|\vec k|$. In this regime the emitted pair is almost collinear
to the photon $k$. Thus the phase space configurations in which $q_3$ and $q_4$
are back-to-back are highly suppressed by $t$-channel dynamics.

\begin{figure}[ht]
\begin{center}
\includegraphics[bb=140 600 254 681,scale=1.]{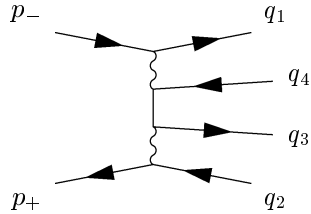}
\end{center}
\caption[multiperipheral]{\label{mul}
                          One of the eight Feynman diagrams for
                          multiperipheral dynamics.}
\end{figure}

However, the selection criteria for kinematic events, used by
the {\sc lep} collaborations and reviewed in section \ref{caloes}, scan also
the hard region. When bremsstrahlung processes get smaller, the next
to leading Feynman graph topology is represented by multiperipheral
dynamics shown in figure \ref{mul}. Notice that this contribution is relevant
also for $\gamma\gamma$ physics, being described in its bulk by
the Weizs\"acker-Williams approximation \cite{WW} for which the internal
photons become quasi-real.

\begin{figure}[ht]
\begin{center}
\includegraphics[bb=140 600 394 681,scale=1.]{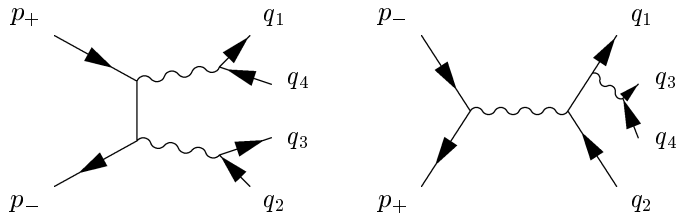}
\end{center}
\caption[extra]{\label{ext}
                Two of the twelve Feynman diagrams representing conversion and
                annihilation dynamics, respectively.}
\end{figure}

Bremsstrahlung and multiperipheral graphs do not complete
all the Feynman graph topologies. Other two classes of diagrams can
be drawn, namely the annihilation and conversion ones, which are shown in
figure \ref{ext}. Their contribution is less important at high
energies and small momentum transfer. Thus in this paper phase space
parameterization and importance sampling does not deal with these configurations.\\

The two following subsections show how the kinematics is treated according to
the previous considerations about the dynamics.

\subsection{Phase Space for Bremsstrahlung Events}

\subsubsection{Phase Space Parameterization}

Let us consider the four-body phase space for two incoming particles of momentum
$p_-$ and $p_+$, and four outgoing particles of momentum $q_1 \dots q_4$.
Let us define $P\equiv p_-+p_+$.

\begin{equation}
\label{phspvel}
d^8R_4(P;q_1,q_2,q_3,q_4) \equiv \prod_{i=1}^4{d^3 q_i \over 2E_i}
\delta^{(4)}(P-\sum_{i=1}^4 q_i)
\end{equation}

The four-body kinematics can be split into the product of simpler
processes. $t$-channel dynamics suggests to think of the event as
formed by a fermionic current interacting with a particle which
radiates a pair. Thus let us deal with the event as a three body
plus a two body kinematics. The following decomposition can be
introduced

\begin{equation}
d^8R_4(P;q_1,q_2,q_3,q_4) = d^5R_3(P;q_1,q_2,q_{34})d^2R_2(q_{34};q_3,q_4)ds_{34}
\end{equation}

\noindent where $s_{34}$ is the squared mass of $q_{34} \equiv q_3+q_4$.

The three body kinematics can be expressed in terms of invariants. Let us define
$s_{ij\dots} \equiv (q_i+q_j+\dots)^2$ and
$t_{\pm j\dots} \equiv (p_\pm-q_j-\dots)^2$.
A straightforward calculation leads to 

\begin{equation}
d^5R_3(P;q_1,q_2,q_{34}) = {1\over 32\lambda^{1/
2}(s,m^2,m^2)\sqrt{-\Delta_4}} \;ds_{12}dt_{-34}ds_{134}dt_{+2}d\varphi
\end{equation}

\noindent where $d\varphi$ integration is over the beam direction, $\lambda$
is the K\"allen function and $\Delta_4$ is the fourth order symmetric Gram
determinant.

The two body kinematics gets the simpler form

\begin{equation}
d^2R_2(q_{34};q_3,q_4) = {\lambda^{1/2}(s_{34},m^2,m^2)\over
8s_{34}}d^2\Omega_3^*
\end{equation}

\noindent where $\Omega_3^*$ is the solid angle of $q_3$ in the frame in which
$\vec q_{34} = 0$.

\subsubsection{Physical Region}

The physical region depends on the choice of the order of integration for the phase
space variables. Let us consider the following choice (read up-down then left-right)

\begin{tabbing}
\hspace*{12em} \= $ds_{34}$\hspace*{1em} \= $dR_3$\hspace*{1em} \= $dR_2$ \\
\> \makebox[8.5em][c]{\hrulefill}\\
\> $s_{34}$  \> $s_{12}$ \> $\cos\theta_3^*$ \\
\>          \> $t_{-34}$ \> $\varphi_3^*$ \\
\>	  \> $s_{134}$ \\
\>	  \> $t_{+2}$
\end{tabbing}

\noindent With this ordering the limits for the phase space variables are

\begin{eqnarray}
s_{34}  & \in & [4m^2,(s^{1/2}-2m)^2] \nonumber \\
s_{12}  & \in & [4m^2,(s^{1/2}-s_{34}^{1/2})^2] \nonumber \\
t_{-34} & \in & \left[m^2+s_{34}-\Big({s+s_{34}-s_{12}\over 2}+
                \sqrt{{1\over 4}-{m^2\over s}}\lambda^{1/2}
                (s,s_{34},s_{12})\Big),\right. \nonumber \\
        &     & \left.,m^2+s_{34}-\Big({s+s_{34}-s_{12}\over 2}-
                \sqrt{{1\over 4}-{m^2\over s}}\lambda^{1/2}
                (s,s_{34},s_{12})\Big)\right] \\
s_{134} & \in & \left[m^2+s_{34}-\Big({s_{12}+s_{34}-s\over 2}+
                \sqrt{{1\over 4}-{m^2\over s_{12}}}\lambda^{1/2}
                (s,s_{34},s_{12})\Big),\right. \nonumber \\
        &     & \left.,m^2+s_{34}-\Big({s_{12}+s_{34}-s\over 2}-
                \sqrt{{1\over 4}-{m^2\over s_{12}}}\lambda^{1/2}
                (s,s_{34},s_{12})\Big)\right] \nonumber \\
                t_{+2}  & \in & \{t_{+2}: \Delta_4 \leq 0 \} \nonumber
\end{eqnarray}

\noindent These limits are exact, thus the phase space generation has efficiency
equal to one if no selection criterion is set on.

\subsubsection{Importance Sampling}
\label{bremwg}

The soft pair limit provides a relatively simple analytical approximation
for the $t$-channel contribution to the cross section.
This expression can be reached either from direct
calculation or from analytic results available in the literature \cite{soft}.
The resulting integral gives a guideline to sample the full cross section formula.

According to the choice for the phase space variables ordering, the weights $w$ follow

\begin{eqnarray}
w(s_{34})  & = & {1\over s_{34}}\nonumber\\
w(s_{12})  & = & \left\{\begin{array}{ll}
		         1/(s+s_{34}-s_{12})\propto(1/E_{34})
                           & ,s_{12}\geq s'_{12}\\
                         1 & ,s_{12}\leq s'_{12}
                        \end{array}\right.\nonumber\\
w(t_{-34}) & = & {1\over m^2-t_{-34}}\\
w(s_{134}) & = & {1\over s_{134}-m^2}\nonumber\\
w(t_{+2})  & = & \left\{\begin{array}{ll}
		         1 & ,t_{+2}\leq t'_{+2}\\
		         1/(t_{+2}^2) & ,t'_{+2}\leq t_{+2}\leq t''_{+2}\\
		         1/(-t_{+2}) & ,t_{+2}\geq t''_{+2}
                        \end{array}\right.\nonumber
\end{eqnarray}

\noindent where $w(s_{34})$ and $w(s_{12})$ deal with the infrared pole,
$w(t_{-34})$ and $w(s_{134})$ with the collinear pole, and
$w(t_{+2})$ with the coulomb pole. The boundaries $s'_{12}$, $t'_{+2}$
and $t''_{+2}$ should be set according to the selection criteria.

\subsection{Phase Space for Multiperipheral Events}

Multiperipheral events require true four-body kinematics, since no natural
expansion of the phase space can fit the dynamics, because of the three
propagators between the ingoing particles. Thus a description based on
energies and angles can be preferred to an invariant picture.

\subsubsection{Phase Space Parameterization}

\noindent Let us integrate out $q_4$ and put $E_i\equiv q_{i,0}$

\begin{equation}
d^8R_4(P;q_1,q_2,q_3,q_4) = \delta(E-\sum_{i=1}^4 E_i){d^3 q_1 \over 2E_1}
{d^3 q_2 \over 2E_2}{d^3 q_3 \over 2E_3}{1\over 2E_4} 
\end{equation}

\noindent where $E_1$, $E_2$ and $E_3$ are given by mass shell relations and
$E_4$ by energy-momentum conservation.

Let us consider the decay $k\rightarrow q_3,q_4$ where
$k\equiv q_3+q_4$, $\omega\equiv k_0$, and $s_{34}\equiv k^2$.
The three momentum $|\vec q_3|$ and the cosine of the polar angle
$\theta_3$, with respect to a fixed direction $\hat {\vec u_z}$,
are needed as phase space variables. 
Let us write $\vec k$ as $\vec k = k_z \hat {\vec u_z} + k_x \hat {\vec u_x}$
the third axis being defined as $\hat {\vec u_y}=\hat {\vec u_z} \times \hat {\vec u_x}$.
Energy conservation implies that

\begin{eqnarray}
\omega^2 + E_3^2 - 2 \omega E_3 &=& m^2 + |\vec q_3|^2 + |\vec k|^2 -2 k_z |\vec q_3| \cos\theta_3 + \nonumber \\
                                & & - 2 k_x |\vec q_3| \sin\theta_3\sin\varphi_3
\end{eqnarray}

\noindent namely

\begin{equation}
\sin\varphi_3 = {1\over 2 k_x |\vec q_3| \sin\theta_3} ( 2 \omega E_3 - s_{34}
-2 k_z |\vec q_3| \cos\theta_3 )
\end{equation}

\noindent The Jacobian is easily obtained as

\begin{equation}
J^{-1} = { k_x |\vec q_3| \sin\theta_3\cos\varphi_3 \over E_4}
\end{equation}

\noindent This leads to the following phase space parameterization

\begin{eqnarray}
d^8R_4(P;q_1,q_2,q_3,q_4) &=& {1\over 16 k_x  E_1 E_2 E_3 |\vec q_3|
                               \sin\theta_3\cos\varphi_3} \times \nonumber \\
                          & & d|\vec q_1| d^2\Omega_1 d|\vec q_2| d^2\Omega_2
                              d|\vec q_3| d\cos\theta_3
\end{eqnarray}

\noindent where $k_x=(\vec P - \vec q_1 - \vec q_2)\cdot \hat {\vec u_x}$,
$E_i = \sqrt{m^2+ |\vec q_i|}$, and $d^2\Omega_i\equiv d\cos\theta_i d\varphi_i$.

\subsubsection{Physical Region}

\noindent The physical region is given by

\begin{equation}
| 2 k_x |\vec q_3| \sin\theta_3 | \ge | 2 \omega E_3 - \omega^2 + s_{34}
-2 k_z |\vec q_3| \cos\theta_3 |
\end{equation}

\noindent which means ($\sin\theta_3\ge 0$)

\begin{equation}
\begin{array}{lll}
2 ( k_x |\vec q_3| \sin\theta_3 - \omega E_3 + k_z |\vec q_3| \cos\theta_3 ) & \ge &
- \omega^2 + s_{34} \\
2 ( k_x |\vec q_3| \sin\theta_3 + \omega E_3 - k_z |\vec q_3| \cos\theta_3 ) & \ge &
\omega^2 - s_{34}
\end{array}
\end{equation}

\noindent By taking into account that $|\vec q_3|\le\sqrt{|\vec q_3|^2+m^2}\le |\vec q_3|+m$
it follows that the range of interest is certainly contained into the domain

\begin{equation}
\begin{array}{lll}
2 ( k_x  \sin\theta_3 - \omega  + k_z  \cos\theta_3 ) |\vec q_3| & \ge &
- \omega^2 + s_{34} \\
2 ( k_x  \sin\theta_3 + \omega  - k_z  \cos\theta_3 ) |\vec q_3| & \ge &
\omega^2 - s_{34} -2 \omega m
\end{array}
\end{equation}

\noindent which leads to

\begin{equation} 
{  \omega^2 - s_{34} \over 2 ( k_x  \sin\theta_3 + \omega  - k_z  \cos\theta_3 ) }
\le |\vec q_3| \le 
{  \omega^2 - s_{34} \over 2 ( \omega - k_x  \sin\theta_3  - k_z  \cos\theta_3 ) }
\end{equation}

\noindent if  $\omega - k_x  \sin\theta_3  - k_z  \cos\theta_3 > 0$

\noindent The maximum and minimum of $|\vec q_3|$ are achieved for $\vec q_3$ collinear to
$\vec k$ in the forward and backward direction, respectively.

\subsubsection{Importance Sampling}
\label{multiwg}

The Weizs\"acker-Williams equivalent photon approximation \cite{WW} provides
a useful guideline to deal with the contribution of multiperipheral
diagrams. The resulting integral shows how to sample the full cross section
formula by introducing the weights

\begin{equation}
w(\cos\theta_1) = {1\over -t_{-1}(\cos\theta_1)} \quad,\quad
w(\cos\theta_2) = {1\over -t_{+2}(\cos\theta_2)}
\end{equation}

\noindent where $w(\cos\theta_1)$ and $w(\cos\theta_2)$ mimic the most singular
behaviour of the Weiz\-s\"ac\-ker-Williams spectrum. $t_{-1}$ and $t_{+2}$ are
equal to $(p_- - q_1)^2$ and $(p_+ - q_2)^2$, respectively, as previously
defined. Further a flat importance sampling can be made on $\cos\theta_3$
to match the selection criteria better.

\section{\label{caloes} Selection Criteria}

Two types of selection criteria are considered: {\tt BARE1} and {\tt CALO2}.
They were defined by the "Event Generators for Bhabha Scattering" working
group during the {\sc cern} Workshop "Physics at {\sc lep2}" (1994/1995),
see ref.~\cite{common} for more details.

These algorithms are tailored to Bhabha scattering with photon
emission. Thus they must be modified to fit with pair emission, chiefly
to deal with identical particles and with fermion clusters.

\subsection{A Simple Setup: {\tt BARE1}}

This non-calorimetric criterion selects events with one or more particles per
calo. If more than one particle hits a calo it chooses the most energetic
hit, the reference particle. Thus for each kinematic event two particles
are labelled as the emitted pair and the other two as the beam particles.

\begin{table}[ht]
\caption[acceptance]{\label{angle}
                     Angular acceptances in degrees for {\tt BARE1}
                     and {\tt CALO2} algorithms.}
\medskip
\begin{center}
\begin{tabular}{|c||c|c||c|c|} \hline
   & \multicolumn{2}{c||}{\tt BARE1}        & \multicolumn{2}{c|}{\tt CALO2} \\
\cline{2-5}
   & Left Calo           & Right Calo          & Left Calo           & Right Calo \\
\hline\hline
WW & $2.70\rightarrow7.00$ & $2.70\rightarrow7.00$ & $2.97\rightarrow6.73$ & $2.97\rightarrow6.73$ \\
\hline
NN & $3.30\rightarrow6.30$ & $3.30\rightarrow6.30$ & $3.49\rightarrow6.11$ & $3.49\rightarrow6.11$ \\
\hline
NW & $3.30\rightarrow6.30$ & $2.70\rightarrow7.00$ & $3.49\rightarrow6.11$ & $2.97\rightarrow6.73$ \\
\hline
\end{tabular}
\end{center}
\end{table}

Then the beam particles are selected by their energies and angles.
The angular cut can be made with the same acceptance for the two caloes
(wide-wide and narrow-narrow), or with different acceptances (narrow-wide).
\begin{equation}
   \begin{array}{lll}
      3.3^\circ \leq\theta\leq 6.3^\circ &-& {\rm narrow\;acceptance}\\
      2.7^\circ \leq\theta\leq 7.0^\circ &-& {\rm wide\;acceptance}
   \end{array}
\end{equation}
\noindent  Angular acceptances are summarized in table \ref{angle}.
Further the energy cut is imposed by considering
the adimensional variable $z$ as usually done in the literature
\cite{jadach}
\begin{equation}
\label{zed}
      z \equiv 1-{E_{\rm 1}E_{\rm 2}\over E^2_{\rm beam}} \leq z_{\rm max}
\end{equation}

It is worth noticing that small values of $z$ means soft pair
emission, while large values allows for hard pair radiation.

\subsection{A Calorimetric Setup: {\tt CALO2}}

This selection criterion looks for any particle in two caloes,
the former is along the ingoing
electron direction, the latter is along the ingoing positron direction.
Then it scans the detected particles in each calo to pick up the most
energetic one, the reference particle.

\begin{figure}[ht]
\begin{center}
\includegraphics[scale=.8]{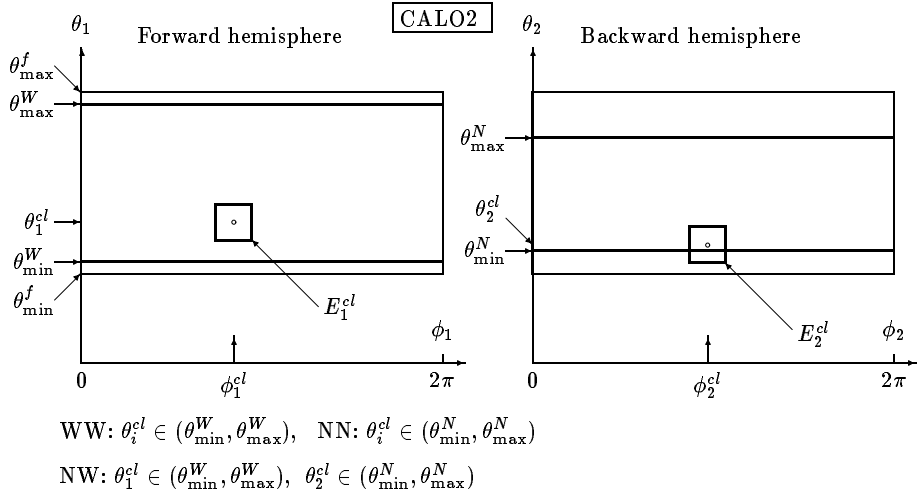}
\end{center}
\caption[calo2]{\label{calo2}
               Geometry and acceptances of {\tt CALO2} calorimetric setup.
               From ref.~\cite{common}.}
\end{figure}

It defines a cluster for each reference particle. The shape of the cluster
is a square in the $(\theta,\varphi)$ plane of the corresponding calo, its
size is of $3\Delta\theta/16 \times 3\pi/16$ radiants, where
$\Delta\theta$ is the calo polar width.
The cluster center is pinned to the reference particle. The cluster energy
is the sum of the energies of the particles in the cluster itself.

The last step of the algorithm is to reject the events according to
an angular and an energy cut, where angles are referred to the cluster center
and energies to the cluster energy.
Each polar width must be reduced to an effective acceptance,
as shown in figure \ref{calo2}, to prevent the cluster square
to exceed the calo size, i.e. should the reference particle be out
of the effective acceptance but within the calo total width, yet the
event is rejected.

The angular cut can be made with the same polar width for the two
caloes, defining a wide-wide and a narrow-narrow acceptance, or
with different widths for the two caloes, defining a narrow-wide
acceptance
\begin{equation}
   \begin{array}{lll}
      3.3^\circ \leq\theta\leq 6.3^\circ &-& {\rm width\; for\; narrow\;acceptance}\\
      2.7^\circ \leq\theta\leq 7.0^\circ &-& {\rm width\; for\; wide\;acceptance}
   \end{array}
\end{equation}
\noindent  Angular acceptances are summarized in table \ref{angle}.
Further the energy cut is imposed by considering the adimensional
variable $z$ as given by eq.~(\ref{zed}).

\section{\label{algdyn} Dynamics Calculation}

Generators including the full set, or part, of the {\sc qed} diagrams
for $e^+e^-\rightarrow e^+e^-l^+l^-$ ($l=e,\mu,\tau$)
was already described in ref.~\cite{bdk} and they are used for analysis and
simulation of $\gamma\gamma$ collision processes at {\sc lep} and other
$e^+e^-$ colliders.

In the present paper the {\sc mc} program {\tt PAIRS}, written in {\tt FORTRAN},
was built to compute the $e^+e^-\rightarrow e^+e^-e^+e^-$ process.
It implements the importance samplings and the selection criteria sketched
previously, and computes the exact matrix element, including all
mass terms, by using the {\tt ALPHA} algorithm and the resulting code
\cite{alpha}, that is conceived for the automatic computation of tree-level
multi-particle production amplitudes without any need of
Feynman graphs expansion.

Without entering the details of the algorithm {\tt ALPHA} it is worth
noticing, for the aim of the present study, that the predictions
of this automatic algorithm have been already compared with
diagrammatic results for processes with four fermions in the final
state \cite{fferm} showing an excellent agreement, and also
successfully used to obtain original results for other reactions
\cite{photons}.

The main features of the program {\tt PAIRS} can be summarized
as follows. The code computes the phase space integral by means
of an importance sampling both for bremsstrahlung and
multiperipheral graphs. Since the contribution of the other
topologies is small, as discussed in section \ref{phsp}, there is
no need of a specific strategy to reduce the associated variance.

The integration is performed in two steps. In the former the matrix
element is sampled by using the bremsstrahlung weights, see
section \ref{bremwg}. Then a rejection algorithm selects the events
belonging to the phase space region $\Omega_B$ which is preferred by the
bremsstrahlung dynamics. In the latter the same calculation scheme
is performed in terms of the multiperipheral weights, see
section \ref{multiwg}, and of a rejection algorithm selecting
the remaining phase space region $\Omega_M$, which is preferred
by the multiperipheral dynamics. The treatment of the interference
between the two dynamics is addressed below.

After phase space generation the program deals with identical particles and
other symmetries of the amplitude. Feynman expansion in pure {\sc qed}
shows sixteen bremsstrahlung graphs, eight with initial-state emission
and eight with final-state emission of a fermion pair.
These two families can be sampled at the same time
leading only to eight different weights, which can
be obtained, fixed one of them, by repeated application of
identical fermion exchange symmetry ($ID$ symmetry)
or of $CP$ symmetry. Let us note that under these symmetries
the squared amplitude does not change. There are also eight multiperipheral
graphs, four with an amplitude value and four, obtained twisting
the kernel fermion lines ($TW$ symmetry), with a different value.
The two multiperipheral squared amplitudes are left invariant
by the $ID$ symmetry.

The symmetries of the integrand allow us to cast the cross section
formula in a simpler form, because the phase space jacobians and
the selection criterion characteristic functions share these symmetries too.

An example can clarify the importance of the integrand symmetries. Let us
consider a function $f(x,y)$, symmetric under the change $x \rightarrow y$,
which is integrated over the unit square in the $(x,y)$ plane.
In order to draw the calculation some weights are introduced, for definiteness 
say $\omega_1\equiv 1/(x+a)$ ($a>0$) is a suited weight because
of some pathology in $x=-a$, but, since $f(x,y)$ is symmetric under coordinate
exchange, it must also exhibit the same pathology in $y=-a$. Thus the weight
$\omega_2\equiv 1/(y+a)$ must be introduced too.
The integration by using importance sampling follows.
\begin{eqnarray}
 & & \int_0^1 \,d x \int_0^1 \,d y \; f(x,y) =
     \int_0^1 \,d x \int_0^1 \,d y \; \left( \frac{\omega_1 f(x,y)}{\omega_1+\omega_2}+\frac{\omega_2 f(x,y)}{\omega_1+\omega_2} \right) = \nonumber \\
 &=& \;\int_0^1 \, \frac{dx}{x+a} \int_0^1 \,d y \; \frac{(x+a)(y+a)}{x+y+2a}f(x,y) + \nonumber \\
 & & + \int_0^1 \,d x \int_0^1 \, \frac{dy}{y+a} \; \frac{(x+a)(y+a)}{x+y+2a}f(x,y) = \nonumber \\
 &=& \; 2 \int_0^1 \,d x \int_{\ln a}^{\ln (a+1)} \,d \mu \; \frac{(x+a)(y+a)}{x+y+2a}f(x,y)
\end{eqnarray}
\noindent In the last step coordinate exchange symmetry both of $f(x,y)$ and
integration region was used, and the integration measure changed in order to include
the weight by putting $d\mu\equiv dy/(y+a)$.

Hence the illustrated two channel sampling can be collapsed into a single integration,
because the integrand and the integration region share the same symmetries. The
same strategy can also be adopted for the more complex integral involved in the present
calculation.

Let us name by $x$ the phase space coordinates, by
$d^8R_4(x)$ the phase space volume element defined by eq.~(\ref{phspvel}),
by $N$ the cross section normalization,
by $\chi(x)$ the characteristic function of
the selection criterion, by $p^B_i(x)$ the eight sets of weights for
the bremsstrahlung amplitude and by $p^M_i(x)$ and  $p^M_i(x_{\rm TW})$
the four weights for each of the multiperipheral amplitudes.
These quantities share the symmetry properties

\begin{equation}
\begin{array}{llll}
d^8R_4(x_i)    &=& d^8R_4(x)      & \forall i \in \{ID,CP,TW\}\\
\chi(x_i)      &=& \chi(x)        & \forall i \in \{ID,CP,TW\} \\
|M(x_i)|^2     &=& |M(x)|^2       & \forall i \in \{ID,CP\} \\
p^{B,M}_i(x_j) &=& p^{B,M}_j(x_i) & \forall i,j \in \{ID,CP,TW\}
\end{array}
\end{equation}

\noindent where $x_i$ is a phase space point obtained from $x$ applying
a suited symmetry among $ID$, $CP$, and $TW$.

\noindent In terms of these quantities the cross section integral becomes

\begin{eqnarray}
\label{master1}
\sigma &=& 8N\int_{\Omega^B}\,{d^8R_4(x)p^B(x)\chi(x)\over\sum_{i\in\{ID,CP\}}p^B_i(x)}
           \sum_{\rm spin}|M(x)|^2 + \nonumber \\
       & & 4N\int_{\Omega^M}\,{d^8R_4(x)p^M(x)\chi(x)\over\sum_{i\in\{ID,TW\}}p^M_i(x)}
           \sum_{\rm spin}\left(|M(x)|^2+|M(x_{TW})|^2\right)
\end{eqnarray}

\noindent This technique is very useful because it permits to treat twenty-four
channels as they were three.

The above division of phase space leaves some problem in the
bremsstrahlung region because of the interferences between
$t$-channel and multiperipheral dynamics. Thus it is useful
to add an extra flat channel to the first integral of eq.~(\ref{master1}).
The flat channel deals also with a multiperipheral
contribution so it must be sampled further by the $p^M_i(x)$ weights.
Hence the cross section integral is

\begin{eqnarray}
\label{master2}
\sigma &=& 8N\int_{\Omega^B}\,{d^8R_4(x)p^B(x)\chi(x)\over\sum_{i\in\{ID,CP\}}p^B_i(x)+\eta}
           \sum_{\rm spin}|M(x)|^2 + \nonumber \\
       & & 8N\eta\int_{\Omega^B}\,{d^8R_4(x)p^M(x)\chi(x)\over p^M(x)\left(
           \sum_{i\in\{ID,CP\}}p^B_i(x)+\eta\right)}
           \sum_{\rm spin}|M(x)|^2 + \nonumber \\
       & & 4N\int_{\Omega^M}\,{d^8R_4(x)p^M(x)\chi(x)\over\sum_{i\in\{ID,TW\}}p^M_i(x)}
           \sum_{\rm spin}\left(|M(x)|^2+|M(x_{TW})|^2\right)
\end{eqnarray}

\noindent where $\eta$ is a suitable weight for the flat channel.

The matrix element is computed by functionally integrating the {\sc qed}
tree level effective lagrangian with the iterative algorithm {\tt ALPHA},
which gives directly the total amplitude $M(x)$.

After the cross section is computed for the real contribution, also the virtual
correction \cite{burgers} corresponding to pair emission is added. The resulting
pair contribution is then normalized against the tree level Bhabha
scattering cross section. The final result can be corrected for initial-state
radiation via collinear structure functions too \cite{alta}.

\section{\label{ancomp} Comparison between {\sc mc} Results and Analytical Calculations}

Before the numerical results being worked out, the {\sc mc}
program is now tested against analytical results already present
in the literature \cite{russi}. Such calculations are available for fixed
selection criteria, among which one resembling {\tt BARE1} algorithm.

\begin{figure}[ht]
\begin{center}
\includegraphics[bb=55 385 280 610,scale=.8]{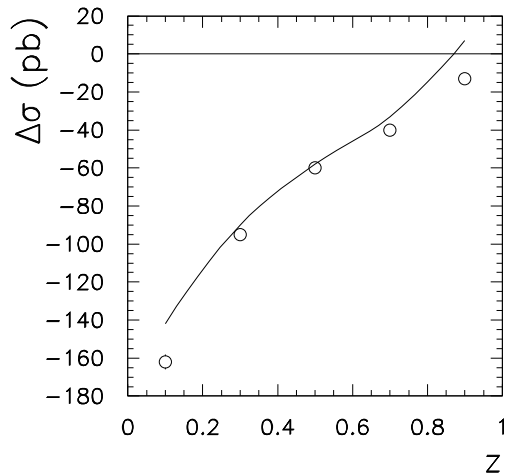}
\end{center}
\caption[analytical]{\label{rus}
                     Comparison between {\sc mc} integration of the exact
                     matrix element
                     (markers) and analytical computation of ref.~\cite{russi}
                     (solid line) as a function of the energy cut $z$, as given
                     by eq.(\ref{zed}). Entry values
                     sum up real plus virtual cross sections, and are computed
                     for {\tt BARE1} setup with symmetric acceptance
                     $1.375^\circ\leq\theta\leq 3.323^\circ$ at
		     $\sqrt{s}=92.3{\rm~GeV}$.}
\end{figure}

A direct comparison, which could be affected by some bias because of
analytical approximations and possibly small differences in the selection
criterion, is shown in figure \ref{rus} and table \ref{rust} for the absolute
value of the pair correction. Solid line represents analytical
computation of ref.~\cite{russi}, while markers show the results of the
present {\sc mc} calculation. The error bars are within
the markers. Entry values are in $\mathrm{pb}$, sum up both real and
virtual part, and are computed for the {\tt BARE1}
setup with symmetric angular acceptance $1.375^\circ\leq\theta\leq 3.323^\circ$
at $\sqrt{s}=92.3{\rm~GeV}$.

\begin{table}[ht]
\caption[analyticalt]{\label{rust}
                      Comparison between present numerical results and analytical
                      ones of ref.~\cite{russi} with the same settings of figure \ref{rus}.
                      Entry values are in $\mathrm{pb}$. In this setup Bhabha
                      tree level cross section is $\sigma_0=175426(14) \;\mathrm{pb}$.
                      The error on $\sigma_0$ is due to numerical integration.}
\medskip
\begin{center}
\begin{tabular}{|l||c|c|c|c|c|} \hline
           & $z=0.1$      & $z=0.3$     & $z=0.5$     & $z=0.7$     & $z=0.9$ \\ \hline \hline
Present {\sc mc} & $-162\pm 4$ & $-95\pm 1$ & $-60\pm 2$ & $-40\pm 3$ & $-13\pm 3$ \\ \hline
Analytical & $-142$      & $-90$      & $-58$      & $-33$      & $ +7$ \\ \hline
\end{tabular}
\end{center}
\end{table}

\noindent Table \ref{rust} shows that the relative difference between
numerical and analytical results is within $15\%$ for the experimentally
relevant values of $z$ \break ($0.3\leq z\leq0.7$), thus providing a rather satisfactory
test of the {\sc mc} program. Further checks were performed with the results of
ref.~\cite{russi} for {\tt CALO2} event selection leading to a similar behaviour.

\section{\label{tchn} Comparison between Exact Matrix Element and $t$-Channel Approximation}

Numerical computations already present in the literature are limited
to bremsstrahlung graphs without fermion exchange and up-down
interference \cite{jadach}. Let us define such an approach as $t$-channel
approximation. This approximation, together with the analytical work of
ref.~\cite{russi}, is an essential ingredient to estimate
the theoretical error associated with light pair correction to {\sc sabh}
scattering.

\begin{figure}[ht]
\begin{center}
\includegraphics[bb=49 230 550 555,scale=.75]{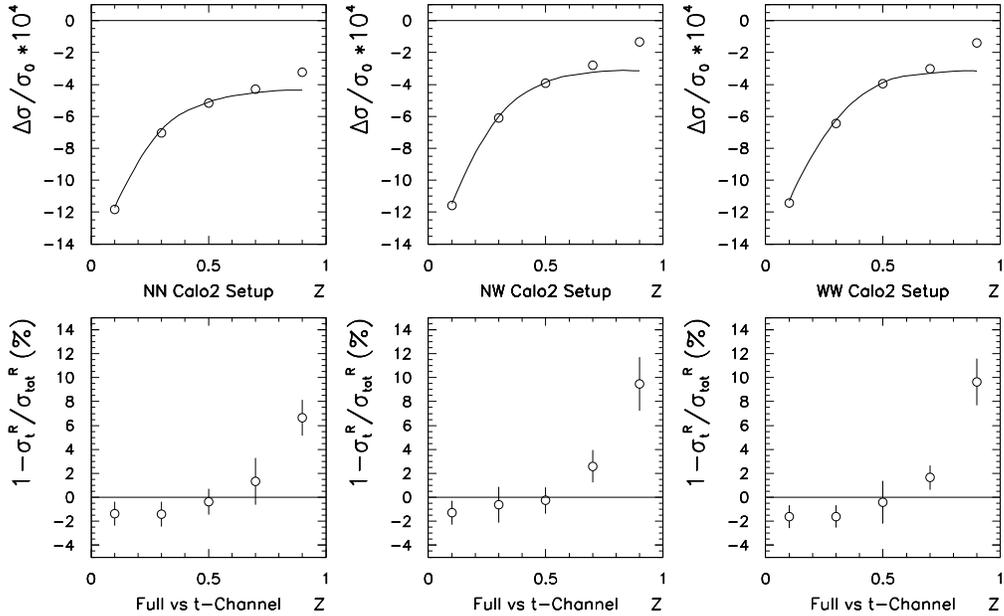}
\end{center}
\caption[vary]{\label{vary}
               In the upper row: exact matrix element (markers)
               vs $t$-channel approximation (solid line) as a function
               of the energy cut $z$, as given by eq.(\ref{zed}). Entry values
               sum up real plus virtual cross sections, and are computed
               by varying the angular acceptances of {\tt CALO2} setup
	       at {\sc lep1/slc} energies.
               In the lower row: relative difference between the exact matrix
               element and $t$-channel approximation still as a function
               of the energy cut. Entry values refer only to the real
               pair production cross section.}
\end{figure}

As far as the {\sc mc} calculation of ref.~\cite{jadach} is concerned,
the uncertainty is the sum of a physical error, due to the incomplete
matrix element calculation, and a technical error,
due to algorithm stability and finite {\sc cpu} time.
In ref.~\cite{jadach} the physical error associated to the matrix element
for real pair production is estimated to be $30\%$. Once the virtual
correction is added, the cancellation against the real part can increase
the magnitude of the error. In ref.~\cite{jadach} summing up all the sources of 
physical and technical uncertainty leads to a conservative estimate of
the light pair correction of $-1.3\times10^{-4} \pm 2\times10^{-4}$
for narrow-wide {\sc calo2} acceptance at $z=0.5$ at {\sc lep1/slc} energies.
Thus the comparison of $t$-channel approximation with the full matrix
element computation can size the magnitude of the physical error, and
therefore lowers the theoretical error.

In order to remove any spurious effect in the comparison between
such calculations and the present one, the amplitude has been
directly calculated in $t$-channel approximation by using
the {\tt ALPHA} algorithm\footnotemark
\footnotetext{Since the ALPHA approach does not make use of Feynman 
graph expansion, this is not entirely straightforward. The goal
is achieved as follows: the input lagrangian is modified by introducing
two distinct photons $\gamma_1$ and $\gamma_2$. The electron couples
to both photons, whereas the muon and the tau lepton couples
to $\gamma_1$ and $\gamma_2$, respectively. Within this modified
lagrangian $m_\tau=m_\mu=m_e$ is assigned and the process
$e^+ \mu^- \rightarrow e^+ \mu^- \tau^+ \tau^-$ is studied. It is immediately
seen that this reproduces the $t$-channel approximation.}
and comparing that amplitude with the full amplitude within the same
selection criteria. Further the same virtual correction
\cite{burgers} for both the cross sections is implemented.
The results are shown in figure \ref{vary} for {\tt CALO2}
setup and different angular acceptances at {\sc lep1/slc} energies.
In the first row the solid line represents the $t$-channel approximation,
while the markers the results of the present full calculation.
Entries are normalized to Bhabha
tree level cross section ($\sigma_0$) and sum up both real and
virtual part. In the second row the markers represent the relative difference
between the real part of the $t$-channel approximation 
($\sigma_t^{\rm R}$) and the real part of the full calculation
($\sigma_{\rm tot}^{\rm R}$).

The comparison shows appreciable differences only for $z\geq 0.7$,
without much sensitivity to angular acceptance variation.
By increasing $z$ greater than $0.7$, the relative difference between
the real part of the cross sections can grow up to $15\%$.

\begin{table}[ht]
\caption[finalt]{\label{finalt}
                 Comparison between exact and $t$-channel approximation
                 numerical results for narrow-wide {\tt CALO2} setup, see figure \ref{final}.
                 Entry values are in $\mathrm{pb}$ except for the last column.
                 In this setup Bhabha tree level cross section
                 is $\sigma_0=21939(1) \;\mathrm{pb}$.
		 The error on $\sigma_0$ is due to numerical integration.}
\medskip
\begin{center}
\begin{tabular}{|c|r|r||r|r|} \hline
       & Full Dynamics      & $t$-Channel       & Abs.Diff.        & Rel.Diff. ($\%$) \\ \hline \hline
$z=0.1$ & $-25.36 \pm 0.01$ & $-25.48 \pm 0.02$ & $ 0.12 \pm 0.02$ & $ 0.48 \pm 0.08$ \\ \hline
$z=0.3$ & $-12.85 \pm 0.05$ & $-13.34 \pm 0.02$ & $ 0.49 \pm 0.05$ & $ 3.67 \pm 0.38$ \\ \hline 
$z=0.5$ & $ -7.14 \pm 0.05$ & $ -8.43 \pm 0.02$ & $ 1.29 \pm 0.05$ & $15.30 \pm 0.63$ \\ \hline
$z=0.7$ & $ -4.98 \pm 0.12$ & $ -7.06 \pm 0.06$ & $ 2.08 \pm 0.14$ & $29.46 \pm 2.23$ \\ \hline
$z=0.9$ & $ -1.75 \pm 0.17$ & $ -6.90 \pm 0.07$ & $ 5.15 \pm 0.19$ & $74.64 \pm 3.51$ \\ \hline
\end{tabular}
\end{center}
\end{table}

\begin{figure}[ht]
\begin{center}
\includegraphics[bb=55 385 280 610,scale=.8]{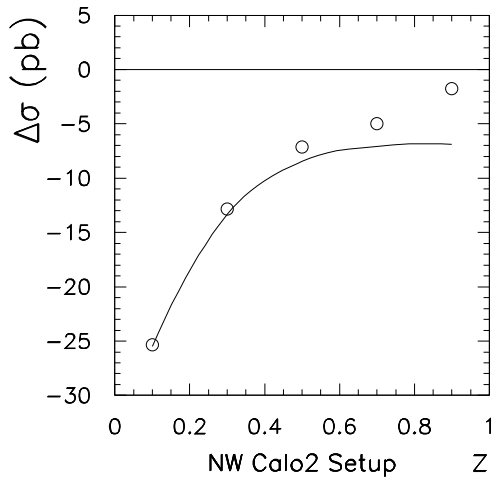}
\end{center}
\caption[final]{\label{final}
                Results of a refined computation of the exact matrix element (markers)
                vs $t$-channel approximation (solid line) as a function
                of the energy cut $z$, as given by eq.(\ref{zed}). Entry values
                sum up real plus virtual cross sections, and are computed
                for narrow-wide {\tt CALO2} setup.}
\end{figure}

A more refined computation is shown in figure \ref{final},
allowing a higher statistics, to get an improved accuracy in order to
establish a better evaluation of the physical error. The result is shown
in figure \ref{final} and table \ref{finalt} for {\tt CALO2} narrow-wide
setup. The solid line represents $t$-channel approximation, while the
markers the full calculation. The error bars are within the markers.
Entry values are in $\mathrm{pb}$, and sum up both real and virtual
part. Table \ref{finalt} shows the numerical values for the two
calculations and the absolute and relative difference. Entry
values are in $\mathrm{pb}$.

The relative difference between the $t$-channel approximation and
the full calculation listed in table \ref{finalt} shows that
the physical error, introduced by using the $t$-channel approximation,
is within $30\%$ for the relevant values of $z$.
In the soft region (small $z$ values) the relative difference gets
smaller since the bremsstrahlung diagrams are the largely dominating
contribution at this regime. On the other hand, in the hard region (large $z$ values)
the multiperipheral diagrams become more and more relevant, worsening
the agreement with the $t$-channel approximation.

Some tests performed at {\sc lep2} energies show the physical error behaviour
given in table \ref{lep2t}. The difference between the exact calculation and
the $t$-channel approximation is within $10\%$ for the experimentally relevant
$z$ values.

\begin{table}[ht]
\caption[lep2]{\label{lep2t}
               Comparison between exact and $t$-channel approximation numerical results
               for narrow-wide {\tt CALO2} setup at {\sc lep2} energies.
               Entry values are in $\mathrm{pb}$ except for the last column.
               In this setup Bhabha tree level cross section
               is $\sigma_0=6087.0(3) \;\mathrm{pb}$.
	       The error on $\sigma_0$ is due to numerical integration.}
\medskip
\begin{center}
\begin{tabular}{|c|r|r||r|r|} \hline
        & Full Dynamics    & $t$-Channel      & Abs.Diff.        & Rel.Diff. ($\%$) \\ \hline \hline
$z=0.3$ & $-4.47 \pm 0.01$ & $-4.27 \pm 0.01$ & $-0.20 \pm 0.02$ & $-4.61 \pm 0.24$ \\ \hline 
$z=0.5$ & $-2.91 \pm 0.01$ & $-2.68 \pm 0.01$ & $-0.23 \pm 0.02$ & $-8.70 \pm 0.62$ \\ \hline
$z=0.7$ & $-2.43 \pm 0.01$ & $-2.28 \pm 0.01$ & $-0.15 \pm 0.02$ & $-6.62 \pm 0.64$ \\ \hline
\end{tabular}
\end{center}
\end{table}

The previous results refer to the {\tt CALO2} calorimetric setup
explained in section~\ref{caloes}, where the calo acceptances
are set according to ref.~\cite{jadach}. Yet some tests were drawn also
with different angular acceptances to match the recent angular standard
of {\sc lep} luminometers, as previously considered in refs.~\cite{common,ward}.
The new ranges are $(1.50,3.20)$ degrees for the wide acceptance and $(1.62,2.84)$
degrees for the narrow acceptance at {\sc lep1/slc} energies. The results are
shown in table \ref{newcalo} confirming the same behaviour previously obtained.

\begin{table}[ht]
\caption[newcalo]{\label{newcalo}
                 Comparison between exact and $t$-channel approximation numerical results
                 for narrow-wide {\tt CALO2} setup at {\sc lep1/slc} energies with a
	         different choice for the angular acceptances: wide $(1.50^\circ,3.20^\circ)$,
                 narrow $(1.62^\circ,2.84^\circ)$. Entry values are in $\mathrm{pb}$
                 except for the last column.
                 In this setup Bhabha tree level cross section
                 is $\sigma_0=87203(4) \;\mathrm{pb}$.
		 The error on $\sigma_0$ is due to numerical integration.}
\medskip
\begin{center}
\begin{tabular}{|c|r|r||r|r|} \hline
       & Full Dynamics      & $t$-Channel       & Abs.Diff.        & Rel.Diff. ($\%$) \\ \hline \hline
$z=0.3$ & $-41.01 \pm 0.30$ & $-43.62 \pm 0.20$ & $ 2.61 \pm 0.36$ & $ 5.99 \pm 0.85$ \\ \hline 
$z=0.5$ & $-23.91 \pm 0.42$ & $-28.91 \pm 0.25$ & $ 5.31 \pm 0.49$ & $18.36 \pm 1.84$ \\ \hline
$z=0.7$ & $-16.99 \pm 0.69$ & $-24.99 \pm 0.25$ & $ 8.00 \pm 0.74$ & $32.02 \pm 3.27$ \\ \hline
\end{tabular}
\end{center}
\end{table}

It must be emphasized that now a {\sc mc} program, that includes
all the {\sc qed} Feynman diagrams for $e^+e^-\rightarrow e^+e^-l^+l^-$ 
($l=e,\mu,\tau$), is available to calculate pair production
corrections. Hence the physical error for electron real pair
production can be reduced well below the $30\%$ quoted
in the literature. Further it is worth noticing that muon pair production
was checked to be one order of magnitude smaller than the electron one.

\section{\label{isrcorr} Initial-State Radiation}

Corrections by {\sc isr} can be evaluated by the aid of collinear {\sc qed}
structure functions in the non-singlet approximation
\cite{alta}. Numerical computations for the previously discussed
$t$-channel approximation are known in the literature \cite{jadach}.
Moreover, corrections of the order of $O(\alpha^3L^3)$, as due to
single bremsstrahlung to pair production, was analytically calculated
in ref.~\cite{russi}.

\begin{figure}[ht]
\begin{center}
\includegraphics[bb=35 305 535 475,scale=.75]{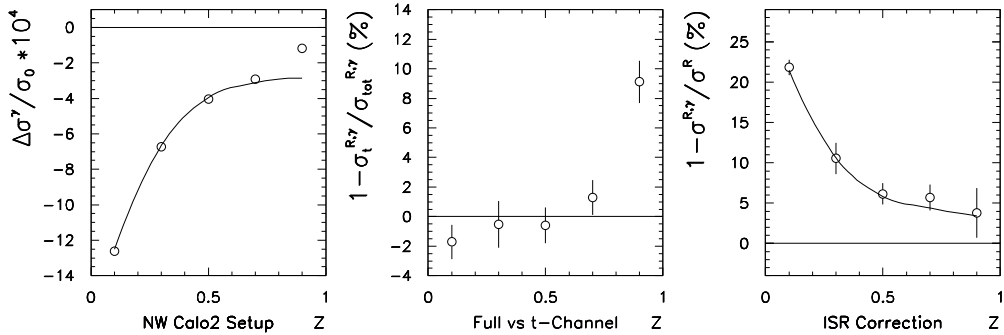}
\end{center}
\caption[isr]{\label{isr}
               First picture: exact matrix element (markers)
               vs $t$-channel approximation (solid line) as a function
               of the energy cut $z$, as given by eq.(\ref{zed}), in the presence
               of {\sc isr}. Entry values sum up real plus virtual cross sections,
               and are computed for narrow-wide angular acceptances of {\tt CALO2} setup
	       at {\sc lep1/slc} energies.
               Second picture: relative difference between the exact matrix
               element and $t$-channel approximation till as a function
               of the energy cut in the presence of {\sc isr}. Entry values
               refer only to the real pair production cross section.
               Third picture: relative effect of {\sc isr} correction for exact matrix
               element and $t$-channel approximation.}
\end{figure}

In figure \ref{isr} the comparison between the exact matrix element and
$t$-channel approximation is made in the presence of {\sc isr} with the {\tt CALO2}
setup in the narrow-wide acceptance at {\sc lep1/slc} energies and with the same notation
previously used. An apex $\gamma$ means that the relative correction is dressed
by {\sc isr}.

As can been seen, the presence of {\sc isr} becomes more and more important in the soft region
(small $z$ values) up to modifying pair correction by a $25\%$, but it does not
alter the analysis in tree level approximation discussed in the previous section.

\section{\label{end} Conclusions}

The progress in the reduction of the sources of theoretical error
to Bhabha cross section at small angle, driven by increased
experimental accuracy, showed the need of reducing the uncertainty 
associated to the light pair contribution.

Till now the theoretical error to {\sc sabh} scattering due to pair
production is evaluated by using {\sc mc} results based on $t$-channel approximation
and with the aid of analytical means. In order to improve the present situation,
a new {\sc mc} program for $e^+e^- \rightarrow e^+e^-e^+e^-$ process
was presented. It includes the exact {\sc qed} four-fermion matrix element, {\sc isr} in
the collinear approximation, and realistic selection criteria.
The results obtained by means of that program were compared with
the best approximations present in the literature, either analytical or numerical.

Two kinds of contribution form the theoretical error: the physical error,
due to matrix element approximation, and the technical error, due to
numerical integration. The comparison between the exact matrix element 
calculation performed in the present paper and the $t$-channel approximation
adopted in the literature showed that the physical error committed
by using such approximation at {\sc lep1/slc} energies is within $30\%$
with the exception of the hard region. Indeed in such region it grows up
to $75\%$ of the correction, while in the soft and intermediate region it
can be much lower. Thus the calculation drawn with the $t$-channel approximation
is within the physical error quoted in the literature \cite{jadach}
in the experimentally relevant kinematic region. At {\sc lep2} energies
the physical error due to the approximation is smaller and it is within $10\%$.

Concerning the present calculation, the physical error associated to electron pairs
amounts in neglecting $Z$-boson contributions to the tree level matrix element,
in approximating {\sc isr} by the collinear structure functions,
and in omitting pair correction exponentiation as for instance done in
ref.~\cite{jadach}. Relatively to the whole correction, the physical error
due to these sources can be estimated within $5\%$, 
while the technical error amounts only to a few per cent, leading to
a conservative $10\%$ of the correction for the total theoretical error.

Further light pair production sums up the electron pair
production with other contributions, e.g. muon pair production. Such
effect amounts for muon pair production to a $10$-$20\%$ of
the electron contribution. However, since the present algorithm can
compute also the effect of muon pairs, the total theoretical error
can be estimated at $30\%$ level of the whole correction, as due to 
approximations in electron and muon pair corrections and to neglecting
heavier pair contributions.

A review for the contribution to the theoretical error at {\sc lep1/slc}
is shown in table \ref{error}. In the first column are listed
the values of table \ref{sabs} \cite{common} and in the second column 
the updated values to the latest results for photonic $O(\alpha^2L)$
corrections \cite{ward,pv,kr}, while in the third column the
values are further updated to the result of the present analysis, when
taking into account electron and muon pair corrections and a total
error estimate of $30\%$.

\begin{table}[ht]
\caption[error]{\label{error}
                Theoretical error in {\sc sabh} scattering at {\sc lep1/slc}.
                The first column (see also table \ref{sabs}) refers to ref.~\cite{common},
                the second column takes into account the results of refs.~\cite{ward,pv,kr},
                the third column updates to the present analysis.}
\medskip
\begin{center}
\begin{tabular}{|l||c|c||c|} \hline
Type of correction/error & ref.~\cite{common} ($\%$) & ref.~\cite{ward} ($\%$) & updated ($\%$) \\ \hline \hline
missing photonic $O(\alpha^2L)$   & $0.100 $        & $0.027 $      & $0.027$ \\
missing photonic $O(\alpha^3L^3)$ & $0.015 $        & $0.015 $      & $0.015$ \\
vacuum polarization               & $0.040 $        & $0.040 $      & $0.040$ \\
light pairs                       & $0.030 $        & $0.030 $      & $0.010$ \\
$Z$-exchange                      & $0.015 $        & $0.015 $      & $0.015$ \\ \hline
total                             & $0.110 $        & $0.061 $      & $0.054$ \\ \hline
\end{tabular}
\end{center}
\end{table}

Table \ref{error} shows that the theoretical error for {\sc sabh}
scattering, because of the reduction of the pair production error,
is now close to $0.05\%$ and so comparable with the experimental accuracy.
The conclusions, drawn for {\sc lep1/slc}, can also be extended to {\sc lep2}
energies and lead to the reduction of the theoretical error shown in table
\ref{error2}.

\begin{table}[ht]
\caption[error2]{\label{error2}
                 Theoretical error in {\sc sabh} scattering at {\sc lep2}.
                 The first column (see also table \ref{sabs}) refers to ref.~\cite{common},
                 the second column takes into account the results of refs.~\cite{ward,pv,kr},
                 the third column updates to the present analysis.}
\medskip
\begin{center}
\begin{tabular}{|l||c|c||c|} \hline
Type of correction/error & ref.~\cite{common} ($\%$) & ref.~\cite{ward} ($\%$) & updated ($\%$) \\ \hline \hline
missing photonic $O(\alpha^2L)$   & $0.200 $        & $0.040 $      & $0.040$ \\
missing photonic $O(\alpha^3L^3)$ & $0.030 $        & $0.030 $      & $0.030$ \\
vacuum polarization               & $0.100 $        & $0.100 $      & $0.100$ \\
light pairs                       & $0.050 $        & $0.050 $      & $0.015$ \\
$Z$-exchange                      & $0.000 $        & $0.000 $      & $0.000$ \\ \hline
total                             & $0.250 $        & $0.122 $      & $0.113$ \\ \hline
\end{tabular}
\end{center}
\end{table}

It is worth noticing that now the major contribution to the theoretical
error is due to the hadronic uncertainty in the vacuum polarization correction. Thus,
if the new determinations \cite{vacuum} of $\alpha_{\rm QED}(M_Z)$
could be drawn to the $1{\rm\;GeV}$ mass scale too, the total error could decrease
further.

The present work was conceived to lower the theoretical error on
{\sc sabh} scattering due to light pair production.
Yet, in order to manage the numerical and theoretical questions,
techniques of a wider range of applicability were implemented.
Chiefly, to deal with multiperipheral dynamics, Weizs\"acker-Williams
approximation played an important r\^ole as a guideline to develop a suited
importance sampling and it led to a numerical recipe that can be further
applied to perform phenomenological studies of processes involving
particles lost in the beam pipe.

Therefore, future developments could deal with the evaluation of pair correction
to two fermion production large angle processes at present and future $e^+e^-$
colliders. Experimentally relevant processes are $e^+e^-\rightarrow e^+e^-$, whose
interest is for luminosity measurements at {\sc da$\Phi$ne} and {\sc nlc}, and
$e^+e^-\rightarrow$~hadrons in the context of precision studies of the electroweak
interaction at {\sc lep} and beyond.

Other fields of interest concern background processes for $WW$ physics and searches
for new physics, as very recently addressed in ref.~\cite{giamp} for the single $W$
production case, and two photon collision processes.

All these studies will require the upgrade of the {\sc qed} matrix elements considered in the
present paper to the full electroweak ones. It is worth noticing that this generalization
could be worked out quite directly by virtue of the automatic algorithm employed in the
present approach.

\renewcommand{\thesection}{$\!\!\!\!\!\!\!$} \section{Acknowledgements}

M.Moretti was supported by a Marie Curie fellowship ({\sc tmr-erbfmbict} 971934).
A.Pallavicini wishes to thank {\sc infn} Sezione di Pavia for
the support and for computer facilities.

\end{document}